\documentclass{PoS}
                                        
\usepackage{amsmath,amsfonts,enumerate}
    
\newcommand{\ZZ}{\mathbb{Z}}

\newcommand{\CC}{\mathbb{C}}

\title{Continuum limit of the axial anomaly and index for
the staggered overlap Dirac operator: An overview}

\ShortTitle{Axial anomaly and index for the staggered overlap Dirac operator}

\author{David H.~Adams
\\
Division of Mathematical Sciences,        
Nanyang Technological University, Singapore 637371\\
        E-mail: \email{dhadams@ntu.edu.sg}}

\author{\speaker{Reetabrata Har}
\\
Division of Mathematical Sciences,        
Nanyang Technological University, Singapore 637371\\
        E-mail: \email{REET0002@e.ntu.edu.sg}}

\author{Yiyang Jia
\\
Division of Mathematical Sciences,        
Nanyang Technological University, Singapore 637371\\
        E-mail: \email{yyjiahbar@gmail.com}}

\author{Christian Zielinski
\\
Division of Mathematical Sciences,        
Nanyang Technological University, Singapore 637371\\
        E-mail: \email{zielinski@pmail.ntu.edu.sg}}

\abstract{
Evaluation of the continuum limit of the axial anomaly and index is sketched for 
the staggered overlap Dirac operator. There are new complications compared to the usual 
overlap case due to the distribution of the spin and flavor 
components around lattice hypercubes in the staggered formalism.
The index is found to correctly reproduce the continuum index, 
but for the axial anomaly this is only true after averaging over the sites of a lattice 
hypercube. 
}

\FullConference{31st International Symposium on Lattice Field Theory - LATTICE 2013\\
		July 29 - August 3, 2013\\
		Mainz, Germany}

\begin{document}

\section{Introduction}

Overlap fermions, as a massless lattice fermion formulation, have exact zero-modes with
definite chirality and hence a well-defined index \cite{Neu(overlap-GW)}.
Furthermore, they satisfy \cite{Neu(overlap-GW)} the Ginsparg-Wilson (GW) relation 
\cite{GW-Hasenfratz} and therefore have an exact chiral symmetry which can be identified with
the axial U(1) symmetry of continuum massless Dirac fermions \cite{Luscher(GW)}. 
The latter is broken by quantum effects -- specifically by the fermion measure \'a la
Fujikawa \cite{Fuji(continuum-anomaly)} -- so there is an axial anomaly. 
It is proportional to an index density for the index of the overlap Dirac operator, just
like in the continuum.
(The continuum anomaly is proportional to the 
topological charge density, whose integral -- the topological charge of the background
gauge field -- is equal to the index by the Index Theorem.)
An important test of the overlap lattice fermion 
formulation is then to show that the continuum anomaly and index
are reproduced in the continuum limit in smooth gauge field backgrounds. This has been 
verified in a number of papers at various levels of rigor 
\cite{GW-Hasenfratz,KY-Fuji(lattice-anomaly)-Suzuki,DA(AnnPhys),DA(JMP)}, and also in a more
general setting where the kernel operator in the overlap formulation is a more general 
hypercubic lattice Dirac operator \cite{DA-Biet}.

The advantageous theoretical properties of overlap fermions are offset by the high cost
of implementing them in numerical simulations of lattice QCD.
Recently, a staggered version of the overlap fermions, describing 2 fermion species 
(flavors), was introduced in \cite{DA(PLB)} as a further development of the spectral flow 
approach to the staggered fermion index in \cite{DA(PRL)}. The fermion field 
in this case is a one-component field (no spinor indices) just like for staggered fermions.
This offers the prospect of theoretical advantages of overlap fermions at a cheaper 
cost.\footnote{An exploratory investigation of the cost of staggered overlap fermions
was made in \cite{Forcrand}. More recently, the cost of staggered Wilson fermions was 
investigated in \cite{DA(lattice13)}; the results there give positive indications for the
cost effectiveness of staggered overlap fermions on larger lattices.} 

To establish a secure theoretical foundation for staggered overlap fermions, one of the 
things that needs to be done is to verify that it reproduces the continuum anomaly and index
in the continuum limit in smooth gauge field backgrounds, just like for usual overlap 
fermions. The purpose of this paper is to sketch the verification of this. Full details
will be given in a later article.\footnote{Numerical checks that the staggered overlap 
Dirac operator has the correct index in smooth backgrounds have already been done in
\cite{DA-Andriy} (for lattice transcripts of instanton gauge fields) and in \cite{Forcrand}.} 
There are new complications for evaluating the continuum limit of the anomaly and index
in this case compared to the usual overlap case, due to the fact that the 
spin and flavor components are distributed around the lattice hypercubes in the staggered 
formalism. 
It turns out that, although the index correctly reproduces the continuum index, the 
axial anomaly only reproduces the continuum anomaly after averaging over the sites of a 
lattice hypercube. This is not surprising, since the sites around a lattice hypercube
can be regarded as corresponding to the same spacetime point in the staggered 
formalism.  

We focus on the original 2-flavor version of staggered overlap fermions introduced in 
\cite{DA(PLB)}. A 1-flavor version of staggered Wilson fermions was later introduced in
\cite{Hoel} and can be used as kernel to make a 1-flavor version of staggered overlap
fermions. However, this formulation has the drawback of breaking lattice rotation symmetry;
a new gluonic counterterm then arises which needs to be included in the bare action
and fine-tuned to reproduce continuum QCD, thus reducing the attractiveness of this 
formulation for practical use. Nevertheless, it correctly reproduces the continuum 
axial anomaly and index; this can be shown by a straightforward modification of our arguments
here for the 2-flavor case; the details will be given in the later article.

\section{The staggered overlap Dirac operator, its index, and the axial anomaly}

The staggered overlap Dirac operator $D_{so}$ is given by
\begin{equation}
D_{so}=\frac{1}{a}\Big(1+A/\sqrt{A^{\dagger}A}\Big)\ ,\quad A=D_{sW}-\frac{m}{a}
\label{1}
\end{equation}
where $D_{sW}$ is the staggered Wilson Dirac operator, obtained by adding a ``Wilson term''
to the staggered fermion action as discussed below (see (\ref{9})).
The staggered Wilson term reduces the number of fermion species (flavors)
described by the staggered fermion from 4 to 2 (the other 2 species get masses 
$\sim1/a$ and become doublers) \cite{DA(PLB)}. 
Consequently, for suitable choice of $m$ in (\ref{1}), the staggered overlap fermion describes
2 physical fermion flavors. Specifically, the requirement is $0<m<2$ just like for usual 
overlap fermions.

The role of unflavored $\gamma_5$ in this setting is played not by the unflavored $\gamma_5$
of the staggered fermion but by the {\em flavored} $\gamma_5$ which gives the exact flavored
chiral symmetry of the staggered fermion action, which we denote by $\Gamma_{55}$.
This notation reflects the fact that it
corresponds to $\gamma_5\otimes\xi_5$ in the spin-flavor interpretation of
staggered fermions by Golterman and Smit \cite{GS}, where $\{\xi_{\mu}\}$ is a representation 
of the Dirac algebra in flavor space.\footnote{$\Gamma_{55}$ is denoted by $\epsilon$ in 
\cite{GS}.}  
The use of $\Gamma_{55}$ as the unflavored $\gamma_5$
in the staggered versions of Wilson, domain wall and overlap fermions is justified by the 
fact that it acts in a unflavored way on the physical fermion species, since $\xi_5=1$ on 
these \cite{DA(PLB)}. 

With $\Gamma_{55}$ playing the role of unflavored $\gamma_5$, the requirements 
$\gamma_5^2=1$ and $\gamma_5$ hermiticity of the staggered Wilson Dirac operator 
$D_{sW}$ are satisfied, and hence the staggered overlap Dirac operator satisfies the 
the Ginsparg-Wilson (GW) relation \cite{GW-Hasenfratz}, which in this case is
\begin{equation}
\Gamma_{55}D_{so}+D_{so}\Gamma_{55}=aD_{so}\Gamma_{55}D_{so}.
\label{2}
\end{equation}
Moreover, (\ref{1}) can be rewritten as
\begin{equation}
D_{so}=\frac{1}{a}\Big(1+\Gamma_{55}H/\sqrt{H^2}\Big)
\label{3}
\end{equation}
where
\begin{equation}
H=H_{sW}(m)=\Gamma_{55}(D_{sW}-m/a)
\label{4}
\end{equation}
is a hermitian operator. 

The index of $D_{so}$ is then determined by the spectral flow of $H_{sW}(m)$ just like in the
usual overlap case, and can be computed from the index formula in the 2nd paper of
\cite{GW-Hasenfratz}:
\begin{equation}
\mbox{index}\,D_{so}=-\frac{1}{2}\mbox{Tr}\Big(H/\sqrt{H^2}\Big).
\label{5}
\end{equation}
Here Tr is the operator trace for operators on the space of lattice staggered fermion fields,
i.e. functions living on the lattice sites and taking values in the fundamental representation
of the color gauge group SU(3) (color indices but no spinor indices). 

The lattice spacetime is the 
usual hypercubic discretization of a 4-dimensional box of fixed length $L$, with $N$ sites
along each axis so that the lattice spacing is $a=L/N$. The link variable associated with a 
lattice link $[x,\,x+a\mu]$ is denoted $U_{\mu}(x)$ or $U(x,x+a\mu)$. (The index $\mu$ is also
used to denote the unit vector along the positive $\mu$-axis.) 
The link variables are taken to be the lattice transcripts of a smooth continuum gauge field
$A_{\mu}(x)$ taking values in the Lie algebra of SU(3). Boundary conditions are imposed 
as in \cite{DA(JMP)} by requiring fields at opposite ends of the box to be related by gauge 
transformations in such a way that $A_{\mu}(x)$ can be topologically nontrivial with topological 
charge $Q\in\ZZ$ (see \cite{DA(JMP)} for the details). The box contains finitely many lattice
sites, so the vector space of lattice fermion fields is finite-dimensional and hence
the index (\ref{5}) is well-defined for all choices of $m$ for which $H_{sW}(m)$ does not 
have zero-modes. In the following we restrict for simplicity to the physical choices
where $0<m<2$.

From (\ref{5}) the index can be expressed as a sum over lattice sites 
(``lattice spacetime integral'') of a density $q(x)\,$:
\begin{equation}
\mbox{index}\,D_{so}=\sum_x a^4\,q(x)\ ,\quad 
q(x)=-\frac{1}{2}\mbox{tr}_c\Big(\frac{H}{\sqrt{H^2}}\Big)(x,x)
\label{6}
\end{equation}
where $\mbox{tr}_c$ denotes the trace over color indices. Here we are using the concept
of operator density ${\cal O}(x,y)$, defined by
${\cal O}\chi(x)=\sum_y a^4\,{\cal O}(x,y)\chi(y)$, which can be used to express the operator
trace as $\mbox{Tr}\,{\cal O}=\sum_x a^4\,\mbox{tr}_c\,{\cal O}(x,x)$.
 
The index density $q(x)$ is seen to be closely related to the axial anomaly in the same
was as for usual overlap fermions \cite{Luscher(GW)}: 
The staggered overlap action $\bar{\chi}D_{so}\chi$ is invariant under the axial 
U(1) transformations $\delta\chi=\Gamma_{55}(1-aD_{so})\chi$, 
$\delta\bar{\chi}=\bar{\chi}\Gamma_{55}$. The corresponding axial current
fails to be conserved: its divergence fails to vanish, and is found in the massless limit to
be ${\cal A}(x)=2iq(x)$, which is by definition the anomaly.

In the usual overlap case,
the main step in showing that both the index and axial anomaly have the correct continuum 
limits is to show that the index density $q(x)$ reproduces the continuum topological charge 
density \cite{DA(AnnPhys),DA(JMP)}:
\begin{equation}
\lim_{a\to0}\;q(x)=-\frac{1}{32\pi^2}\,\epsilon_{\mu\nu\sigma\rho}
\mbox{tr}_c(F_{\mu\nu}(x)F_{\sigma\rho}(x))
\label{7}
\end{equation}
Here $x$ is fixed point in the spacetime box, and the $a\to0$ limit is taken by repeated
subdivisions of the lattice such that $x$ continues to be 
a lattice site of all the subdivided lattices.
However, in the present case, it turns out that (\ref{7}) only holds after $q(x)$ is
averaged over the sites of a lattice hypercube containing $x$. Note that the sites of the 
hypercube all converge to $x$ in the continuum limit, so this situation is not 
unreasonable, and is moreover not surprising in the staggered formalism as
mentioned in the Introduction. 

Our goal in the remainder of this paper is to sketch the derivation of 
the modified version of (\ref{7}), i.e. with $q(x)$ averaged over
the sites of a lattice hypercube containing $x$, and with an extra factor
of 2 due to the two physical fermion flavors described by
the staggered overlap fermion.

A further technical result is required to conclude that the index expression (\ref{6})
converges to the continuum topological charge $2Q$ (which equals the continuum Dirac
index with 2 flavors by the continuum index theorem): 
It needs to be shown that the convergence in (\ref{7}) is uniform in $x$. This can be shown 
in the same way as in the usual overlap case in \cite{DA(JMP)}; we omit the details here.

\section{Continuum limit of the lattice index density}

By the same arguments as in the usual overlap case \cite{DA(JMP)} we have
\begin{equation}
\lim_{a\to0}\;\frac{H}{\sqrt{H^2}}(x,x)
=\lim_{a\to0}\;\int_{-\pi/2a}^{3\pi/2a}d^4p\,e^{-ipx}\,\frac{H}{\sqrt{H^2}}\,e^{ipx}.
\label{8}
\end{equation}
Now decompose the lattice momentum $p\in[-\pi/2a,3\pi/2a]^4$ as
$p=\frac{\pi}{a}A+q$, where $q\in[-\pi/2a,\pi/2a]^4$ and $A=(A_1,A_2,A_3,A_4)$ with 
$A_{\mu}\in\{0,1\}$. In the following, if $A$ and $B$ are two such vectors, we also
consider $A+B$ to be a vector of this type with the components $(A+B)_{\mu}\in\{0,1\}$
mod 2. 
Let $V$ be the vectorspace spanned by the 16 plane waves $\{e_A(x)\equiv e^{i\frac{\pi}{a}Ax}\}$. 
In \cite{GS} two commuting representations of the Dirac algebra are defined on $V$, 
namely $\{\hat{\Gamma}_{\mu}\}$ and $\{\hat{\Xi}_{\nu}\}$, given by
$(\hat{\Gamma}_{\mu})_{AB}=(-1)^{A_{\mu}}\delta_{A,B+\eta_{\mu}}$ and 
$(\hat{\Xi}_{\nu})_{AB}=(-1)^{A_{\mu}}\delta_{A,B+\zeta_{\nu}}$ where $\eta_{\mu}$ and
$\zeta_{\nu}$ are vectors whose components are all zero except $(\eta_{\mu})_{\sigma}=1$ for 
$\sigma<\mu$ and $(\zeta_{\nu})_{\sigma}=1$ for $\sigma>\nu$.\footnote{The hats 
on $\hat{\Gamma}_{\mu}$ and $\hat{\Xi}_{\nu}$ are not
present in the notation of \cite{GS}. We include them here, since we use the unhatted versions
to denote the extensions of these operators from $V$ to the space of one-component 
lattice spinor fields $\chi(x)$.} 
 
The staggered Wilson Dirac operator \cite{DA(PLB)} is $D_{sW}=D_{st}+W_{st}$
where $D_{st}$ is the usual (massless) staggered Dirac operator and 
$W_{st}=\frac{r}{a}(1-\Xi_5)$ is the Wilson term. Here $\Xi_5$ ($=\Gamma_{55}\Gamma_5$) is
an extension of $\hat{\Xi}_5$ from $V$ to the space of staggered lattice fermion fields 
$\chi(x)$, described in \cite{DA(PLB)}.
For present purposes it suffices to note that the action of the staggered Wilson Dirac 
operator on a plane wave $e^{ipx}$ can be expressed as follows. Let $T_{+\mu}$ denote
the parallel transporter given by $T_{+\mu}\chi(x)=U(x,x+a\mu)\chi(x+a\mu)$, and
let $T_{-\mu}$ denote its inverse. The symmetrized covariant derivative is 
$\nabla_{\mu}=\frac{1}{2}(T_{+\mu}-T_{-\mu})$, and we define 
$C_{\mu}=\frac{1}{2}(T_{+\mu}+T_{-\mu})$. We will also need the symmetrized product of the 
$C_{\mu}$s, namely $C_5\equiv\frac{1}{4!}\sum C_{\mu}C_{\nu}C_{\sigma}C_{\rho}$ where the
sum is over all permutations $\{\mu,\nu,\sigma,\rho\}$ of $\{1,2,3,4\}$. 
Then, writing the plane wave as $e^{ipx}=e_B(x)e^{iqx}$, the action of $D_{sW}$ on it 
can be expressed as
\begin{eqnarray}
D_{sW}(e_B(x)e^{iqx})
&=&e_A(x)\Big((\hat{\Gamma}_{\mu})_{AB}\nabla_{\mu}
+\frac{r}{a}\Big(\delta_{AB}-(\hat{\Xi}_5)_{AB}C_5\Big)\Big)\,e^{iqx} \nonumber \\
&=&e_A(x)\Big(\hat{\Gamma}_{\mu}\nabla_{\mu}
+\frac{r}{a}\Big({\bf 1}-\hat{\Xi}_5C_5\Big)\Big)_{AB}\,e^{iqx} 
\;\equiv\;e_A(x)(\tilde{D}_{sW})_{AB}\,e^{iqx}
\label{9}
\end{eqnarray}
Here and in the following there is an implicit sum over repeated indices. This includes the 
vector $A$ which is regarded as an index with 16 possible values.

Since $H$ is built from $D_{sW}$ and $\Gamma_{55}$, and the latter acts on plane waves by
$\Gamma_{55}(e_B(x)e^{iqx})=e_A(x)(\hat{\Gamma}_5\hat{\Xi}_5)_{AB}e^{iqx}$, it follows that 
the action of $H/\sqrt{H^2}$ on a plane wave has the same structure as in (\ref{9}):
\begin{equation}
\frac{H}{\sqrt{H^2}}(e_B(x)e^{iqx})
=e_A(x)\Big(\frac{\tilde{H}}{\sqrt{\tilde{H}^2}}\Big)_{AB}\,e^{iqx}
\label{10}
\end{equation}
where $\tilde{H}$ is obtained from $H$ by replacing $D_{sW}\to\tilde{D}_{sW}$
and $\Gamma_{55}\to\hat{\Gamma}_5\hat{\Xi}_5$.

Using the preceding to evaluate (\ref{8}), we get
\begin{eqnarray}
\lim_{a\to0}\;\frac{H}{\sqrt{H^2}}(x,x)
&=&\lim_{a\to0}\;\sum_B\int_{-\pi/2a}^{\pi/2a}d^4q\,e^{-iqx}e_B(x)\frac{H}{\sqrt{H^2}}\,e_B(x)e^{iqx}
\nonumber \\
&=&\lim_{a\to0}\;\sum_{A,B}e^{i\frac{\pi}{a}(A-B)x}\int_{-\pi/2a}^{\pi/2a}d^4q\,e^{-iqx}\,
\Big(\frac{\tilde{H}}{\sqrt{\tilde{H}^2}}\Big)_{AB}\,e^{iqx}
\label{11}
\end{eqnarray} 
We will show below that (i) the contribution to $q(x)$ from the 
terms in (\ref{11}) with $A=B$ reproduces the continuum topological
charge density (\ref{7}), and (ii) the contributions from the terms with 
$A\ne B$ vanish after averaging $q(x')$ over the sites $x'$ in a lattice hypercube
containing $x$. This will complete the derivation of the averaged version of (\ref{7}).

The sum over $A,B$ with $A=B$ in (\ref{11}) gives
\begin{equation}
\lim_{a\to0}\;\int_{-\pi/2a}^{\pi/2a}d^4q\,e^{-iqx}\,
\mbox{Tr}\Big(\frac{\tilde{H}}{\sqrt{\tilde{H}^2}}\Big)\,e^{iqx}
\label{12}
\end{equation}  
where Tr denotes the trace of a linear operator (matrix) on the vectorspace $V$
spanned by the $e_A(x)$'s.
As shown in \cite{GS}, there is a (non-unique) isomorphism $V\simeq\CC^4\otimes\CC^4$
such that $\hat{\Gamma}_{\mu}\simeq\gamma_{\mu}\otimes{\bf 1}$ and     
$\hat{\Xi}_{\nu}\simeq{\bf 1}\otimes\xi_{\nu}$, where $\{\gamma_{\mu}\}$ and 
$\{\xi_{\nu}\}$ are Dirac matrices on spinor space and flavor space, respectively. 
After choosing a basis for flavor $\CC^4$ such that $\xi_5$ is diagonal, (\ref{9}) gives
\begin{equation}
\tilde{D}_{sW}=(\gamma_{\mu}\otimes{\bf 1})\nabla_{\mu}
+({\bf 1}\otimes{\bf 1})\frac{r}{a}(1\mp C_5)
\label{13}
\end{equation}
with sign $\mp$ for the flavors on which $\xi_5=\pm{\bf 1}$. 
Thus, on each of the two 2-dimensional flavor subspaces on which $\xi_5=\pm{\bf 1}$, 
$\tilde{D}_{sW}$ is a hypercubic lattice Dirac operator of the form considered
in \cite{DA-Biet} (note that $C_5$ couples opposite corners of lattice hypercubes).
Also, $\Gamma_{55}$ becomes $\pm(\gamma_5\otimes{\bf 1})$ in this case.

The free field momentum representation of $C_5$ is 
$C_5(aq)=\prod_{\mu}cos(aq_{\mu})=1+O(a^2)$. Using this, we see from (\ref{13}) that 
$\tilde{D}_{sW}$ describes one physical fermion species for each of the two
flavors of the flavor subspace with $\xi_5={\bf 1}$.
Now note that in this case (\ref{12}) is precisely $\mbox{tr}(H/\sqrt{H^2})(x,x)$ 
for $H=(\gamma_5\otimes{\bf 1})(\tilde{D}_{sW}-m)$. Hence it is given
by the general result of \cite{DA-Biet} on the continuum limit of the anomaly and index
for lattice hypercube fermions.
In (\ref{12}) the part of the trace Tr over the flavor subspace is trivial and just 
produces a factor 2, leaving the trace over spinor space. The general result of 
\cite{DA-Biet} then implies that this contribution of (\ref{12}) 
to $q(x)$ gives the continuum topological charge density (\ref{7}) as claimed, with an extra 
factor of 2 for the two physical fermion flavors.

On the other hand, the contribution of (\ref{12}) coming from the other 2-dimensional
flavor subspace on which $\xi_5=-{\bf 1}$ vanishes. This can also be inferred from the
general result of \cite{DA-Biet}, since in this case the sign $\mp$ is $+$ in 
(\ref{13}) and hence there are no physical fermion species.

It remains to show that the contributions to (\ref{11}) from the terms in the sum over
$A,B$ with $A\ne B$ vanish when averaged over $x\in\{\mbox{a lattice hypercube}\}$.
Writing $x=an$, $n\in\ZZ^4$, we have 
$e^{i\frac{\pi}{a}(A-B)x}=(-1)^{(A-B)n}$.
It is easy to see that summing this over the sites $x$ of a lattice hypercube gives zero
 if at least one of the components of $A-B$ is nonzero (mod 2). 
Consequently, the problem of showing that the hypercube-averaged terms with $A\ne B$ 
in (\ref{11}) vanish is reduced to showing a property of the $x$-dependence of the
integral there, 
\begin{equation}
\int_{-\pi/2a}^{\pi/2a}d^4q\,e^{-iqx}\,
\Big(\frac{\tilde{H}}{\sqrt{\tilde{H}^2}}\Big)_{AB}\,e^{iqx},
\label{15}
\end{equation}   
namely, that this integral changes by $O(a)$ as $x$ is varied among the sites of a lattice 
hypercube. 
The argument for this is as follows. Formally, (\ref{15}) diverges $\sim1/a^4$ for $a\to0$.
However, the integrand can be expanded in powers of the continuum gauge field just like
in the usual overlap case \cite{DA(AnnPhys),DA(JMP)}\footnote{This relies on a certain
bound on the spectrum of the hermitian operator $H$ when the plaquette variable of the 
lattice gauge field satisfies an approximate smoothness condition. It was established
in the usual overlap case in \cite{overlap-local}. An analogous bound
holds in the present staggered overlap case; the proof will be 
given elsewhere.} and it can be shown that only the terms of mass-dimension 4 or higher
in this expansion are nonvanishing.\footnote{This relies on two things: First, exponential
locality of the integrand of (\ref{15}) in the gauge field (shown in the same
way as in the usual overlap case \cite{overlap-local}), which implies
that the terms in the expansion become local functionals of the gauge field in the $a\to0$
limit. Second, the gauge invariance and lattice rotation invariance of staggered overlap 
fermions imply that the local functionals of the continuum gauge field that arise in
the expansion of (\ref{15}) must have mass-dimension $\ge4$. (There are no such terms 
with mass-dimension $\le3$, while for mass-dimension 4 there are the Yang-Mills action
functional and topological charge density.)} For each mass-dimension of the expansion terms
there is an accompanying power of $a$, so the nonvanishing terms contain at least a factor
$a^4$ which balances the divergence $\sim1/a^4$ in (\ref{15}). The fact that (\ref{15})
changes by $O(a)$ as $x$ is varied among neighboring lattice sites then follows from the
fact that the smooth continuum gauge field has this property.  
 
{\bf Acknowledgments.}
D.A. is supported by AcRF grant RG61/10 and a start-up grant from NTU.

\end{document}